\documentstyle[seceqn]{elsart}
\def\ii{{\mathrm{i}}}
\def\bra{\langle}
\def\ket{\rangle}
\def\dint{\int\!\!\!\!\int}
\def\tint{\int\kern-.6em\int\!}
\def\eins{\relax{\rm 1\kern-.25em I}}
\def\coe#1#2{{\textstyle{#1\over #2}}}
\def\frac#1#2{{#1\over#2}}
\def\hyp#1#2{\phantom{\vert}_#1 F_#2}

\def\d{{\mathrm d}}

\input epsf
\def\fig#1#2{\bigskip
   \let\picnaturalsize=N \def\picsize{#1} \def\picfilename{#2}
   \ifx\nopictures Y\else{\ifx\epsfloaded Y\else\fi 
   \global\let\epsfloaded=Y
   \centerline{\ifx\picnaturalsize N\epsfxsize \picsize\fi
   \epsfbox{\picfilename}}}\fi}

\def\gg{{\bf g}} 
\def\Zset{\relax{\hbox{\sf Z\kern-.42em Z}}}\font\sf=cmss10
\def\Rset{\relax{\rm I\kern-.18em R}}

\begin{document}
\begin{frontmatter}
\title{Coulomb Gas on the Half Plane\thanksref{DOE}}
\thanks[DOE]{This work was supported in part by the DOE 
   (DE-FG03-84ER40168).}

\author[USC]{J. Schulze\thanksref{DAAD}}
\address[USC]{Department of Physics, University of Southern California,
   Los\nobreak Angeles,\nobreak CA\nobreak 90089--0484, USA}
\thanks[DAAD]{Supported by the DAAD (Doktorandenstipendium HSPII/AUFE).}

\begin{abstract}
The Coulomb-gas description of minimal models is considered on the
half plane. Screening prescriptions are developed by the perturbative
expansion of the Liouville theory with imaginary coupling and with
Neumann boundary condition on the bosonic field. To generate the
conformal blocks of more general boundary conditions, we propose the
insertion of boundary operations.
\end{abstract}

\end{frontmatter}


\section{Introduction}

The investigation of boundary quantum integrable models
\cite{ghza,war}\ is motivated both by the abundance of applications,
such as the Kondo effect \cite{aflu}, open string theory and quantum
wires \cite{fls}, as well as by the clear description of their bulk
counterparts as perturbations around conformal field theories
\cite{cofl}. In the bulk, Toda field theory \cite{homa,egya}, free
field theory and Coulomb-gas description \cite{dofaA,fel,mat}\ are
most useful tools. In the presence of boundaries, the conformal theory
is well understood \cite{carfour,carnine,beg,buxu}; for systems off
criticality however, integrability sometimes only allows certain
combinations of bulk and boundary terms \cite{bcdr,peza,war}.

The Coulomb gas plays a special role in this game. Being a
(non-affine) Toda field theory, it carries most of the features of the
off-critical systems. On the other hand, it serves perfectly well to
describe minimal conformal models. In this paper, we wish to consider
the Coulomb-gas description of minimal conformal models on the half
plane, and to find connections between its boundary conditions and
screening contours.

In section 2, we will first consider the conformal invariance of the
Toda action. For the half plane, this leads to two conformally
invariant boundary conditions on the free field. One of them is the
Neumann boundary condition, for which the discussion looks most
natural.

Section 3 is a brief review of the Coulomb-gas formulation of minimal
conformal models. The Coulomb gas is treated as a Liouville theory
with imaginary coupling, which is expanded as a marginal perturbation
around a free bosonic field theory. This description with
two-dimensional screening integrals was introduced in \cite{dofaA}\ as
a manifestly conformally invariant alternative. We will compare it to
the contour-integral description, following the lines of Mathur
\cite{mat}\ who considered the full plane.

The Ising model is considered as an example in section 4. We will show
to what contour integrals the two-point functions reduce in the above
mentioned description with Neumann boundary condition on the bosonic
field. In section 4.2, these results are compared to the conformal
blocks of the free or fixed boundary conditions on the spin operator
\cite{carfour,cale}\ which both require boundary-crossing integration 
contours.

This motivates the introduction of new boundary terms in section 5
which are best described as composite operators of a vertex operator
and its mirror image. The expressions introduced are shown to lead
either to vanishing correlators or to sew together screening contours
of the two different half planes, leaving the vertex operator
corresponding to an identity operator of the \mbox{minimal} model at
the boundary. Correspondences to the boundary states in
\cite{carnine}\ are discussed.

The outlook in section 6 compares our boundary expressions to the
boundary terms added to the off-critical action in \cite{ghza}\ and to
the boundary terms in Toda theories \cite{bcdr,peza}. Section 7
summarises the paper.

In the Appendix, it is shown that the two-dimensional integral
fulfills the Ward identities as well for the half plane, and how the
corresponding contour integrals can be derived and evaluated.

\section{Conformal invariance of the Toda action}

Consider, following the approach of \cite{homa}, the Toda action
\begin{equation} \label{action}
   S_{\rm bulk}^{\rm class.}\ =\ \coe1{8\pi} \dint_{M}\!\d^2z\ 
      \sqrt{g}\ \Big[ g^{ab} (\partial_a\Phi)\!\cdot\!(\partial_b\Phi) 
          - \coe2{\beta^2} \sum^r \e^{\beta\,\alpha_i\cdot\Phi} 
          + \coe2\beta R\,\varrho\cdot\Phi \Big]\ ,
\end{equation}
where $g_{ab}$ is the metric of the two-dimensional manifold $M$ with
curvature $R$. The $\alpha_i$ are the simple roots of the ordinary Lie
algebra $\gg$ with rank $r$. Its Weyl vector is $\varrho = \sum^r
\lambda_i$, the fundamental weights $\lambda_i$ are defined by 
$\lambda_i\cdot\alpha_j = \delta_{ij}$. The coupling constant $\beta$
eventually will be sent to $\beta\to \ii \tilde \beta$, with $\tilde
\beta$ real.

The energy-momentum tensor $T_{ab}^{\rm class.} \equiv -\coe{4\pi}
{\sqrt{g}} \frac{\delta S^{\rm class.}}{\delta g^{ab}}$ is traceless
on a flat world sheet ($R=0$) where it simplifies with the help of the
equations of motion to a holomorphic $T(z)$ and an antiholomorphic
$\bar T(\bar z)$:
\begin{eqnarray} 
   T_{zz}^{\rm class.}\ &=&\ -\coe12 (\partial_z\Phi)^2 + \coe1\beta 
      \varrho\cdot(\partial^2_z\Phi)\ \equiv\ T(z)\ ,\nonumber\\
   T_{\bar z\bar z}^{\rm class.}\ &=&\ -\coe12(\partial_{\bar z}\Phi)^2
      +\coe1\beta \varrho\cdot(\partial^2_{\bar z}\Phi)\ \equiv\ 
      \bar T(\bar z)\ ,\nonumber\\
   T_{z\bar z}^{\rm class.}\ &=&\ T_{\bar z z}^{\rm class.}\ =\ 0\ . 
      \label{cemt}
\end{eqnarray}
Under the transformation
\begin{eqnarray} 
   &&g_{ab} \to \Omega^2 g_{ab}\ ,\quad g^{ab} \to \Omega^{-2} g^{ab}\ ,
      \quad \sqrt{g} \to \Omega^2 \sqrt{g}\ , \nonumber\\
   &&\Phi \to \Phi - \coe2\beta \ln \Omega\,\varrho\ ,
      \quad R \to \Omega^{-2} [R - 2 \nabla_a \nabla^a \ln \Omega]\ , 
   \label{ctf}
\end{eqnarray}
the action (\ref{action}) changes to
\begin{eqnarray} 
   S^{\rm class.}\ &\to&\ S^{\rm class.} - \coe1{8\pi} \dint_M\!\d^2z\ 
      \sqrt{g}\ \coe{4\varrho^2}{\beta^2}\ln\Omega\ (R-\nabla_a\nabla^a 
      \ln \Omega)\ - \nonumber\\
   &&\ \phantom{S^{\rm class.}} - \coe1{8\pi} \int_{\partial M}\!\!\!
      \d x\ \sqrt{g}\ \coe4\beta (\partial^\perp \ln \Omega)\ 
      \varrho\cdot(\Phi - \coe1\beta \varrho \ln \Omega)\ , 
      \label{actionA}
\end{eqnarray}
where we have included the possibility of a boundary. The role of the
curvature term $R\,\varrho\cdot\Phi$ in the action (\ref{action}) is
to make the change of the bulk term independent of $\Phi$.

In the absence of boundaries and for vanishing curvature, the action 
is conformally invariant if
\begin{equation} \label{cond}
   \partial_z \partial_{\bar z} \ln \Omega\ =\ 0 \qquad \Rightarrow 
   \qquad \Omega\ =\ g(z)\,h(\bar z)\ ,
\end{equation}
i.e.~conformal invariance restricts to analytic conformal
transformations.  Thus the conformal freedom is equivalent to general
holomorphic changes of $z$, and the model has two independent Virasoro
symmetries, corresponding to $g$ and $h$ in (\ref{cond}). The flatness
of the world sheet is preserved under such transformations.

On the quantum level, normal ordered expressions appear in the action,
the equations of motion, and the energy-momentum tensor \cite{col,man}.  
As a consequence, the energy-momentum tensor (\ref{cemt}) is not 
conserved, and the conformal invariance is broken. To correct this, we 
have to change the prefactor of the curvature term in (\ref{action}) 
from $\coe2\beta$ to $2\,(\beta + \coe1\beta)$. This change does not 
affect the boundary term in (\ref{actionA}), the conformal invariance 
of the action is assured by (\ref{cond}). After having applied the 
quantum equations of motion
\begin{equation} 
   \partial_z\partial_{\bar z}\Phi\ =\ -\coe1\beta \sum^r \alpha_i 
      :\!\e^{\beta\,\alpha_i\cdot\Phi}\!: m^{-2\beta^2}\ ,
\end{equation}
where $m$ is a regulator which eventually can be set to zero, the 
energy-momentum tensor reads \cite{homa}\
\begin{equation} \label{qemt}
   T_{zz}\ =\ -\coe12:\!(\partial_z\Phi)^2\!:\,+\,
      (\beta+\coe1{\beta}\!)\,\varrho\cdot(\partial^2_z\Phi)\ ,
\end{equation}
and similar for $\bar T$. The conformal anomaly of this
energy-momentum tensor is \cite{homa}\
\begin{equation} \label{cano}
   c\ =\ r+12\varrho^2 \Big[ \beta + \coe1\beta \Big]^2\ .
\end{equation}

\subsection{Toda theory on the half plane}

Now consider the same models on the half plane $y>0$, where $z=x+
\ii y$ and the real axis is the boundary $\partial M$, and take into
account the boundary term in (\ref{actionA}), as well. This term still
contains $\Phi$, so it asks for a boundary term in the action which
will cancel the $\Phi$-dependence. One can modify (\ref{action}) to
\begin{equation} \label{actionB}
   S\ =\ S_{\rm bulk} - \coe1{8\pi} \int_{\partial M}\!\!\!\d x\ 
      \coe8{\beta} K \varrho\cdot\Phi\ .
\end{equation}
$K(x)=\Gamma^y_{xx}\vert_{y=0}$ is the extrinsic curvature of the
boundary. The variation of $K$ under (\ref{ctf}) close to the flat
metric is $\delta K = \coe12 (\partial - \bar\partial) \ln\Omega =
-\coe1{2\ii} \partial_y \ln\Omega$. Hence, the change of the boundary
term in (\ref{actionB}) cancels the $\Phi$-dependent boundary term in
(\ref{actionA}), leaving us with the additional $\Phi$-independent
conditions
\begin{equation} \label{bcond}
   \partial_y \ln \Omega \vert_{y=0}\ \equiv\ \ii (\partial - 
      \bar \partial) \ln \Omega \vert_{y=0}\ =\ 0\quad {\rm and}
   \quad K=0\ .
\end{equation}
Since the variation of $K$ is proportional to (\ref{bcond}) the
boundary term in (\ref{actionB}) does not affect the energy-momentum
tensor $T\equiv -\coe{4\pi}{\sqrt{g}}\frac{\delta S}{\delta g}$.

The second condition in (\ref{bcond}) is fulfilled trivially for the
geometry chosen. The first condition in (\ref{bcond}) restricts $g$
and $h$ from (\ref{cond}) to be the analytic continuation of each
other. This implies that the antiholomorphic part of the
energy-momentum tensor has to be the analytic continuation of its
holomorphic part, leaving us with a single copy of the Virasoro
algebra \cite{carfour}. In other words, the component $T_{xy}$ of the
energy-momentum tensor has to vanish on the real axis \cite{car}\
\begin{equation} \label{bvir}
   T_{xy}\ =\ 0\ ,\quad {\rm at}\quad y=0\ .
\end{equation}
Using the normal-ordering procedure of \cite{man}, one finds that
\begin{eqnarray} 
   T_{xy}\ &=&\ \coe1{4\ii} (T_{zz}+T_{\bar z z}-T_{z\bar z}-
      T_{\bar z\bar z})\ =\ \coe1{4\ii} (T_{zz}-T_{\bar z\bar z}) 
      \nonumber\\
   &=&\ \cases{ \qquad\coe18 (\partial_x\Phi)\!\cdot\!(\partial_y\Phi)
         - \coe1{4\beta} \varrho \cdot \partial_x\partial_y\Phi\ 
         &\ classically, \cr         
      \cases{\coe1{16}\Big[ :\!(\partial_x\Phi)\!\cdot\!
            (\partial_y\Phi)\!: + :\!(\partial_y\Phi)\!\cdot\!
            (\partial_x\Phi)\!: \Big] -&\cr
         - \coe14 (\beta+\coe1\beta) \varrho \cdot \partial_x
            \partial_y\Phi&\cr} &\ quantum, \cr}
\end{eqnarray}
and (\ref{bvir}) is fulfilled if one takes the Neumann boundary 
condition
\begin{equation} \label{Neu}
   \partial_y \Phi\vert_{y=0}\ =\ 0\ .
\end{equation}
It can also be fulfilled by taking\footnote{We restrict our
consideration to simply-laced Lie algebras.}
\begin{equation} \label{obc}
   \partial_y \Phi\vert_{y=0}\ =\ \cases{
      \sum \alpha_i \e^{\frac\beta2 \alpha_i\Phi}\ 
         &\ classically \cite{bcdr}, \cr
      \sum \alpha_i :\!\e^{\gamma \alpha_i\Phi}\!:\ m^{-2\gamma^2}
         &\ quantum, \cr}
\end{equation}
where $\gamma=\coe12\beta$ or $\gamma=\coe12\coe1\beta$. Both
conditions (\ref{Neu}) and (\ref{obc}) are conformally invariant,
since -- using that at the boundary $\varphi(z)=\tilde\varphi(\bar z)$
-- the right hand sides of (\ref{obc}) have dimension 1. For the
non-affine $A_1$ (Coulomb gas) which we will consider throughout the
rest of the paper, the two possibilities for $\gamma$ correspond to
the screening operators $V_{\alpha_-}$ and $V_{\alpha_+}$ which are
the natural dimension-one objects one can put on the right hand side
of (\ref{obc}). The same two choices exist as well for all other
simply-laced Lie algebras. For affine algebras, only
$\gamma=\coe12\beta$ has been considered so far \cite{peza}.
 
For the Dirichlet boundary condition $\Phi\vert_{y=0}=0$, the term
$\partial_x\Phi\vert_{y=0}$ would vanish everywhere on the boundary,
but the term $\partial_y\partial_x\Phi$ generically would not. Thus
the Dirichlet boundary condition does not naturally fit into our
description. For the sine-Gordon model, the UV limit of the
appropriate version of (\ref{obc}) is the Neumann boundary condition
(\ref{Neu}), while the IR limit is the Dirichlet boundary condition
$\Phi\vert_{y=0}=\Phi_0$ \cite{ssw}. 

In the following, we will choose Neumann boundary conditions for the
bosonic field, i.e.~we will work at the UV fixed point.

\section{The Coulomb-gas description of minimal models}

After this excursion to Toda field theory, we can try to apply the
results to the Coulomb-gas representation. We will restrict the
discussion to the non-affine algebra $A_1$, and consider the Liouville
theory as a marginal perturbation of a free bosonic field theory. On
the half plane, the bosonic field has to fulfill the Neumann boundary
condition.

\advance\baselineskip by 3pt
Consider the free bosonic field theory ${\cal A} = \coe1{8\pi}
\tint\!\d^2z\ \Big[ (\partial \Phi) (\bar\partial\Phi) + 2\sqrt{2}\ii
\alpha_0 R \Phi\Big]$. Using $r=1$, $\varrho^2=\coe12$ and
$\tilde\beta=-\,\sqrt{\coe{M} {M+1}}$, think of the Liouville
potential in (\ref{action}) as a perturbation
\begin{equation} \label{pert}
   {\cal A}_{\rm pert.}\ =\ \coe1{4\pi\alpha_-^2} \dint\!\d^2z\ 
   :\!\e^{\sqrt{2}\ii\alpha_-\Phi(z,\bar z)}\!:\ . 
\end{equation}

The charges $\alpha_0 = \frac1{\sqrt{4M(M+1)}}$, $\alpha_\pm =
\alpha_0 \pm \sqrt{\alpha_0^2+1}$ and $\alpha_{n,m} = \frac{1-n}2
\alpha_+ + \frac{1-m}2 \alpha_-$ should not be confused with the
simple roots $\alpha_i$ from above. The energy-momentum tensor $T_{zz}
= -\coe12 \nobreak\!:\!\!\nobreak(\partial_z\Phi)^2\!\!:\,+\
\sqrt{2}\ii\alpha_0 \ (\partial^2_z\Phi)$ has the conformal anomaly $c
= 1-24\alpha_0^2 = 1\nobreak-\nobreak\frac6{M(M+1)}$. The vertex
operators $V_{\alpha_{n,m}} = :\nobreak\!\e^{\sqrt{2}\ii\alpha_{n,m}
\varphi}\!:$ carry the charges and conformal weights of the Kac table
\cite{dofaA}.

\advance\baselineskip by -3pt
The perturbation (\ref{pert}) changes the correlation functions to
\begin{eqnarray} 
   &&\bra X \ket_{{\cal A} + {\cal A}_{\rm pert.}}\ = \nonumber\\
   &&=\ \bra \e^{\coe1{4\pi\alpha_-^2}\dint\!\d^2z\ :\e^{\sqrt{2}\ii
      \alpha_-\Phi(z,\bar z)}\!:} X \ket_{\cal A} \nonumber\\
   &&=\ \bra X \ket_{\cal A}\ +\ \coe1{4\pi\alpha_-^2} \dint\!\d^2z\ 
      \bra :\!\e^{\sqrt{2}\ii\alpha_-\Phi(z,\bar z)}\!: X \ket_{\cal A}
      \ +\nonumber\\
   &&\quad\ +\ \coe1{16\pi^2\alpha_-^4} \dint\!\d^2z \dint\!\d^2w\ \bra 
      :\!\e^{\sqrt{2}\ii\alpha_-\Phi(z,\bar z)}\!:\ :\!\e^{\sqrt{2}\ii
      \alpha_-\Phi(w,\bar w)}\!: X \ket_{\cal A}\ +\ \cdots 
      \label{expa}
\end{eqnarray}
for an arbitrary insertion $X$. Charge conservation restricts this
expansion to only one term\footnote{We will abbreviate the coefficient
$\coe1{4\pi\alpha_-^2}$ with $k$. It will later be important, when we
have to find the relative coefficient of conformal blocks with a
different number of screeners. Since we treat the Liouville potential
as a marginal perturbation, $k$ is actually a free parameter which can
be chosen arbitrarily.}. The result is the Coulomb-gas formulation of
minimal models with manifestly monodromy-invariant combinations of the
holomorphic and antiholomorphic sectors. This was introduced in
\cite{dofaB}\ as an alternative to the contour integrals. In
\cite{bnz,mat}, Stoke's theorem was used to show that this description
coincides with the contour-integral description.

The non-vanishing contribution in (\ref{expa}) of the two-point
function with insertion $X = V_{\alpha_{12}} (z_I) V_{\alpha_{12}}
(\bar z_I) V_{\alpha_{12}} (z_{II}) V_{\alpha_{12}} (\bar z_{II})$ is
for example
\begin{eqnarray} 
   &&k \dint\!\d^2z\ \bra V_{\alpha_-}(z) 
      V_{\alpha_-}(\bar z) V_{\alpha_{1,2}}(z_I)V_{\alpha_{1,2}}(\bar z_I) 
      V_{\alpha_{1,2}}(z_{II}) V_{\alpha_{1,2}}(\bar z_{II}) \ket\nonumber\\
   &&=\ k \dint\!\d^2z\ \partial_{\bar z} 
      \int^{\bar z}\!\!\d t\ \bra V_{\alpha_-}(z) V_{\alpha_-}(t) 
      V_{\alpha_{1,2}}(z_I) V_{\alpha_{1,2}}(\bar z_I) 
      V_{\alpha_{1,2}}(z_{II}) V_{\alpha_{1,2}}(\bar z_{II}) \ket \nonumber\\
   &&=\ \coe{-\ii}2 k\!\int_{\partial M}\!\!\d z
      \!\int^{z^*}\!\!\d t\,\bra V_{\alpha_-}(z) V_{\alpha_-}(t) 
      V_{\alpha_{1,2}}(z_I) V_{\alpha_{1,2}}(\bar z_I) 
      V_{\alpha_{1,2}}(z_{II}) V_{\alpha_{1,2}}(\bar z_{II}) \ket\ . 
      \label{trick}
\end{eqnarray}
For the full plane, the integrand splits into a product of a
holomorphic and an anti\-holomorphic factor\footnote{Recall that on
the full plane the two sectors of the free field $\Phi(z,\bar z) =
\varphi(z) + \tilde \varphi(\bar z)$ have a trivial contraction
$\bra\varphi\tilde\varphi\ket=0$, and the vertex operator is
\begin{equation} \label{vert}
   :\!\e^{\sqrt{2}\ii\alpha_-\Phi(z,\bar z)}\!:\ \equiv\ 
   :\!\e^{\sqrt{2}\ii\alpha_-\varphi(z)}\!:\ :\!\e^{\sqrt{2}\ii\alpha_-
   \tilde\varphi(\bar z)}\!:\ .
\end{equation}
Our vertex operators $V_{\alpha_{n,m}}$ and screening operators
$V_{\alpha_\pm}$ are meant to be either the holomorphic or the
antiholomorphic part of such a splitting.}. Mathur \cite{mat}\ treats
the branch cuts as the boundary $\partial M$ along which Stoke's
theorem has to be applied. The $t$-integration splits into contours
between two singular points and the so-called $J$-terms, which go from
a singularity to the complex conjugate $z^*$ of the other integration
variable. The $J$-terms vanish for monodromy reasons. Hence, Mathur is
left with products of holomorphic and antiholomorphic block functions.

For the half plane, the Neumann boundary condition implies the
correlator to be $\bra \varphi(z) \tilde\varphi(\bar w) \ket =
-\ln(z-\bar w)$ and $\bra \tilde\varphi(\bar z) \varphi(w)\ket =
-\ln(\bar z-w)$, and the integrand in (\ref{trick}) will not split
into two sectors\footnote{For the half plane, we have
\begin{eqnarray}
   :\!\e^{\sqrt{2}\ii\alpha_-\Phi(z,\bar z)}\!: 
   &=& (-1)^{-\alpha_-^2}\,:\!\e^{\sqrt{2}\ii\alpha_-\varphi(z)}\!:
      \ :\!\e^{\sqrt{2}\ii\alpha_-\tilde\varphi(\bar z)}\!: \nonumber\\
   &=& (-1)^{+\alpha_-^2}\,:\!\e^{\sqrt{2}\ii\alpha_-\tilde\varphi(\bar z)}\!:
      \ :\!\e^{\sqrt{2}\ii\alpha_-\varphi(z)}\!:\ . \label{foot}
\end{eqnarray}
Reality is ensured by setting $\bra:\!\e^{\sqrt{2}\ii\alpha_-
\Phi(z,\bar z)}\!:\ket = \vert(z-\bar z)^{\alpha_-^2} \vert^2 =
(z-\bar z)^{\alpha_-^2} (\bar z-z)^{\alpha_-^2}$ (and similar), before
replacing $\bar z$ by $t$ in (\ref{trick}).}. The integral
(\ref{trick}) is hence a double integral of two screeners around four
points. This is not surprising since the two-point function on the
half plane has to fulfill the same differential equation as the
holomorphic sector of a four-point function on the full plane
\cite{carfour}. Note, however, that this integral is a double-valued
function. Uniqueness of the full-plane four-point function is obtained
by combining the holomorphic and antiholomorphic blocks in a
monodromy-invariant way. For the half-plane two-point function, both
branches of the double-valued function are invariant under a twist of
the two points in the upper half plane around each other and a
simultaneous twist of their mirror images. The restriction to a unique
function is obtained by observing that the $z$-integration has to stay
in the upper half plane, while the $t$-integration is performed in the
lower half plane.

In a similar way, one has to use for a generic $N$-point function on
the half plane that one cannot get contours which go from one half
plane to the other. Therefore, one gets considerably fewer conformal
blocks than for one sector of the $2N$-point function on the full
plane. Invariance under twists of the points $(z_i-z_j) \to
\e^{2\pi\ii}\,(z_i-z_j)$ and simultaneous twists of their mirror
images $(\bar z_i - \bar z_j) \to \e^{-2\pi\ii}\,(\bar z_i-\bar z_j)$
restricts then to unique results.

\section{Example: The Ising model on the half plane}

In this section and in the Appendix, we want to illustrate what the
boundary $\partial M$ will look like for the half plane, and with what
tools one can evaluate the integral (\ref{trick}).

As an example, consider the Ising model on the half plane. Special
care should be taken of the difference between the Neumann boundary
condition on the free field $\Phi$, which arises naturally in the
Coulomb-gas description, and the free and fixed boundary conditions on
the $\sigma$ operator \cite{carfour}. Throughout the rest of this
paper, it will be understood that the boundary condition on $\Phi$ is
the Neumann boundary condition. Superscripts ``free'' and ``fixed''
will refer to the boundary condition on $\sigma$.

Label the points by $z_1$, $z_2$, $z_3$, $z_4$, \dots\ or by $z_I$,
$\bar z_I$, $z_{II}$, $\bar z_{II}$, \dots\ depending on whether there
is an emphasis on the properties of the holomorphic block on the full
plane or of the $N$-point function on the half plane. See Figure 1 for
an illustration. 

\begin{figure}[h]
\fig{2.2in}{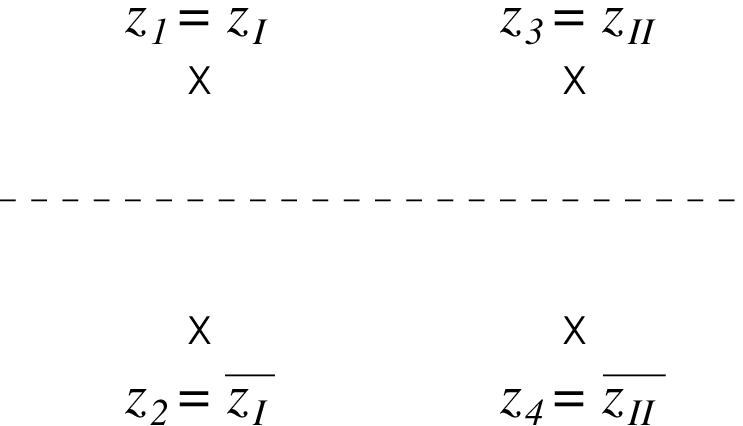}
\caption{Labeling of the points of the two-point function on the 
   half plane}
\end{figure}

Define the cross ratio to be
\begin{equation} \label{cross}
   \xi\ =\ \frac{z_{12}z_{34}}{z_{13}z_{24}} 
      \ =\ -4 \frac{y_I y_{II}}{|z_I-z_{II}|^2} \ ,
\end{equation}
with $z_{ij}=z_i-z_j$ and $z_{I/II}=x_{I/II}+\ii y_{I/II}$. The 
``physical'' cross ratio for the half plane is real negative.

\subsection{Neumann boundary condition on $\Phi$}

In the Appendix, it is shown that for the two-point function of
$V_{\alpha_{1,2}}$, the two-dimensional integral $\tint\d^2z (\cdots)
= \coe{-\ii}2 \int_{\partial M}\d z \int^{z^*}\!\!\!\d t(\cdots)$\
from (\ref{trick}) reduces to one contour in each half plane between
the two respective singularities. From the asymptotic behaviour, this
double integral is seen to be the hypergeometric function
\begin{eqnarray} 
   I_1\ &=&\ \bra V_{\alpha_{1,2}}(z_I) V_{\alpha_{1,2}}(\bar z_I) 
      V_{\alpha_{1,2}}(z_{II}) V_{\alpha_{1,2}}(\bar z_{II}) 
      \ket^{^{\rm Neumann}}_{{\cal A} + {\cal A}_{\rm pert.}}
      \nonumber\\
   &=&\ \coe{-\ii}2 k \int_{z_1}^{z_3}\!\d s \int_{z_2}^{z_4}\!\d t\ 
      \bra \cdots \ket_{\cal A}\nonumber\\      
   &=&\ k_1\ (z_{13}z_{24})^{\frac{2-M}{2(M+1)}}\
      (1-\coe1\xi)^{\frac{2-M}{2(M+1)}}\ \hyp21(\coe{2-M}{M+1},
      \coe1{M+1}; \coe2{M+1}; \coe1\xi)\vert_{M=3}^{\ } \nonumber\\
   &=&\ k_1\ (z_{13}z_{24})^{-\frac18}\ (1-\coe1\xi)^{-\frac18}\
      \coe1{\sqrt{2}} \textstyle{\sqrt{\sqrt{1-\coe1\xi}+1}} \ ,   
      \label{solN}      
\end{eqnarray}
where 
\begin{equation}
   k_1\ =\ k\ B(\coe1{M+1},\coe1{M+1})^2\vert_{M=3}^{\ }\ =\ k\
   B(\coe14,\coe14)^2\ , 
\end{equation}
with $B$ being Euler's Beta function. The expression in the second to
last line is the general result for the two-point function of the
vertex operator $V_{1,2}$ in any minimal model labeled by $M$. For the
determination of the asymptotic behaviour in the Appendix we used that
the integration variables $s$ and $t$ stay in their respective half
planes as illustrated in Figure 2a. The function $I_1$ corresponds to
the conformal block in which the insertions communicate in the
$\eins$-channel through the boundary.

\begin{figure}[h]
\fig{4.6in}{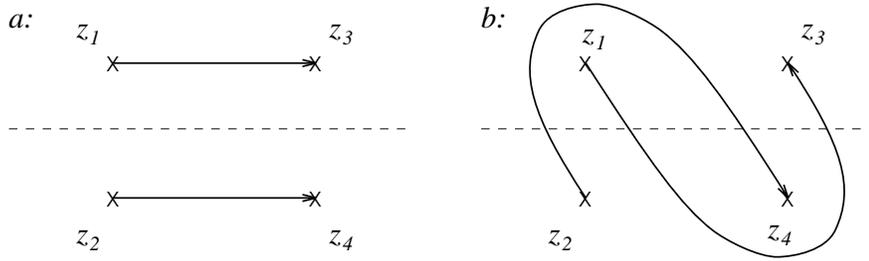}
\caption{\vtop{\hbox{The contours of two conformal blocks of the 
      two-point function.}
   \hbox{a:~Neumann boundary condition on $\Phi$.}
   \hbox{b:~the other conformal block $I_2$.}}}
\end{figure}

Recall that, for the four-point function on the full plane \cite{mat},
this is only one of the conformal block functions of the holomorphic
sector, which consists of any linear combination of (\ref{solN}) with
\begin{eqnarray} 
   I_2\ &=&\ k_2\ (z_{13}z_{24})^{\frac{2-M}{2(M+1)}}\,
      (1-\coe1\xi)^{\frac{2-M}{2(M+1)}}\,(-\coe1\xi)^{\frac{M-1}{M+1}}\,
      \hyp21(\coe1{M+1}, \coe{M}{M+1}; \coe{2M}{M+1}; \coe1\xi)
      \vert_{M=3}^{\ } \nonumber\\
   &=&\ k_1\ (z_{13}z_{24})^{-\frac18}\ (1-\coe1\xi)^{-\frac18}\ 
      \coe1{\sqrt{2}} \textstyle{\sqrt{\sqrt{1-\coe1\xi}-1}}
      \ ,\label{solNb}
\end{eqnarray}
where 
\begin{eqnarray}
   k_2\ &=&\ k\,\coe{M-2}{M-1}\,B(\coe1{M+1},\coe1{M+1})\,
      B(\coe1{M+1},\coe{M-2}{M+1})\vert_{M=3}^{\ } \nonumber\\
   &=&\ \coe{k}2\,B(\coe14,\coe14)^2 \nonumber\\
   &=&\ \coe12\,k_1\ .\
\end{eqnarray}
The function $I_2$ corresponds to the conformal block in which the
insertions communicate in the $\varepsilon$-channel through the
boundary.

The hypergeometric functions in (\ref{solN}) and in (\ref{solNb}) have
a branch cut from $\xi=0$ to $\xi=1$. Crossing this branch cut changes
the hypergeometric function $I_1$ up to a phase $\e^{\frac{2\pi\ii}8}$
to $I_2$, and vice versa. This means that one obtains (\ref{solNb})
from (\ref{solN}) by leaving the ``physical'' region (i.e.~the
negative real axis), analytically continuing over the branch cut and
returning to the ``physical'' region. Therefore, the function
(\ref{solNb}) corresponds (up to the phase $\e^{\frac{2\pi\ii}8}$) to
the contour integral in Figure 2b, in which the variables obviously
leave their respective half planes.

It is important to observe that the two contours in Figure 2b still go
from $z_1$ to $z_3$, and from $z_2$ to $z_4$ respectively. What
changes from one conformal block to the other, is the way these
contours are twisted around the singular points, not the way the
points are paired. The three different ways to pair up the four points
correspond to the expansions around the three singular points of the
cross ratios. An expansion around the zero of a certain cross ratio is
analytic up to the prefactors of type $z_{ij}^{\ \gamma_{i,j}}$ if the
cross ratio tends to zero when bringing the two end points of one
contour close to each other.

Hence, using the Neumann boundary condition, the two-dimensional
screening integrals (\ref{pert}) lead to the conformal block $I_1$,
while the block $I_2$ a priori cannot be obtained by such an integral.

\subsection{The boundary conditions on the spin field $\sigma$}

This subsection summarises the results of \cite{carfour,cale}, and
shows how they fit into the above description. Cardy \cite{carfour}\
uses the facts that for $|z_I-z_{II}| \to \infty$
\begin{equation} 
   \bra\sigma(z_I,\bar z_I)\sigma(z_{II},\bar z_{II})\ket_{\rm h.p.}\
   \to\ \bra\sigma(z_I,\bar z_I)\ket\,\bra\sigma(z_{II},\bar z_{II})\ket\ ,
\end{equation}
and that for the free boundary condition $\bra\sigma(z,\bar
z)\ket^{\rm free} \equiv \bra\sigma(z)\sigma(\bar z)\ket^{\rm free} 
= 0$, even in the limit $y\to0$, i.e.~for $z\to\bar z$ at the
boundary. This leads to the unique solution for the free boundary
condition on $\sigma$
\begin{eqnarray} 
   \bra\sigma(z_I,\bar z_I)\sigma(z_{II},\bar z_{II})\ket_{\rm h.p.}
      ^{\rm free}\
   &=&\ (4 y_Iy_{II})^{-\frac18}\ (1-\xi)^{-\frac18}\ 
      \sqrt{\sqrt{1-\xi}-1} \nonumber\\
   &=&\ \coe1{\sqrt{2}}\,(I_1-I_2)\ , \label{Cardy}
\end{eqnarray}
which is the difference between (\ref{solN}) and (\ref{solNb}), and
has a simple expansion around $\xi=0$, corresponding to the
$\coe1\xi$-expansion of (\ref{solNb})\footnote{The cross ratio $\xi$
in our paper is the inverse of the one used by Cardy \cite{carfour}.}.

On the other hand, in the limit of $z_I$ and $z_{II}$ close to each
other and far away from the boundary, i.e.~for $\xi\to\infty$, one can
use the bulk operator product expansion which does not depend on the
boundary condition\footnote{\cite{beg}\ used the corresponding argument
for the $\Phi_{1,3}$ operator in the $O(N)$ model.}
\begin{equation} 
   \sigma(z_1,z_2)\sigma(z_3,z_4)\ \sim\ \frac{c_1 \eins}
      {(z_{13}z_{24})^{\frac18}} + c_2\,(z_{13}z_{24})^{\frac38} 
      \varepsilon(z_3,z_4) + \cdots\ .
\end{equation}
Parametrising an arbitrary function in the space spanned by
(\ref{solN}) and (\ref{solNb}) as $I = b_1 I_1 + b_2 I_2$, one has
therefore for both the free and fixed boundary conditions 
\begin{eqnarray} 
   b_1\ &=&\ c_1\ =\ \coe1{\sqrt{2}}\ ,\nonumber\\
   b_2\ &=&\ c_2 (-z_{12}z_{34})^{\frac12}\ \bra \varepsilon(z_3,z_4) \ket\ .
      \label{ctwo}
\end{eqnarray}
From equation (\ref{Cardy}), it follows that 
\begin{equation}
   b_2^{\rm free}\ =\ -\coe1{\sqrt{2}}\ . 
\end{equation}
Using equation (\ref{ctwo}), this implies 
\begin{equation}
   c_2\,(-z_{12} z_{34})^{\frac12}\ 
   =\ -\frac1{\sqrt{2}\,\bra\varepsilon(z_3,z_4) \ket^{\rm free}}\ . 
\end{equation}
In \cite{cale}\ it is pointed out that under the duality
transformation interchanging the free and fixed boundary conditions,
the energy operator $\varepsilon$ changes to $-\varepsilon$. Therefore
$\bra \varepsilon(z,\bar z) \ket^{\rm free} = -\bra \varepsilon(z, 
\bar z) \ket^{\rm fixed}$.  Equation (\ref{ctwo}) leads for the fixed
boundary condition on $\sigma$ to
\begin{eqnarray} 
   b_1^{\rm fixed} &=&\ c_1\ =\ \coe1{\sqrt{2}}\ ,\nonumber\\
   b_2^{\rm fixed} 
      &=&\ c_2 (-z_{12}z_{34})^{\frac12} \bra\varepsilon(z_3,z_4)
         \ket^{\rm fixed} 
      \ =\ -\frac{\bra\varepsilon(z_3,z_4)\ket^{\rm fixed}}{\sqrt{2}
         \bra\varepsilon(z_3,z_4)\ket^{\rm free}}
      \ =\ +\coe1{\sqrt{2}}\ . \label{fixed}
\end{eqnarray}
Hence, for fixed boundary conditions, the conformal block is the sum
of the two blocks $I_1$ and $I_2$, and it has as well a clear analytic
behaviour expanding around $\xi=0$, corresponding to the $\coe1
\xi$-expansion of (\ref{solN}).

Equations (\ref{fixed}) and (\ref{Cardy}) being the sum, respectively
the difference, of the two conformal blocks in which the $\eins$- and
$\varepsilon$-channel propagate through the boundary, fits very well
to Cardy's relation between boundary states and boundary conditions
(\ref{carb}) \cite{carnine}.

In terms of contour integrals, this means that the conformal blocks of
the two-point function with free and fixed boundary conditions on
$\sigma$ are given by the integrals $\int_{z_1}^{z_2}\!\d s
\int_{z_3}^{z_4}\!\d t$. For the former, the contours are twisted as
shown in Figure 3a. For the later, they are straight as in Figure
3b. In both cases, one needs boundary-crossing contours which are not
provided by the two-dimensional screening integral (\ref{pert}) using
the Neumann boundary condition.

\begin{figure}[h]
\fig{4.6in}{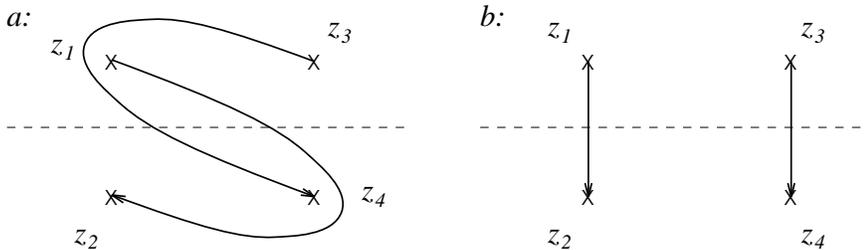}
\caption{\vtop{\hbox{The contours of two other conformal 
      blocks of the two-point function.}
   \hbox{a: free boundary condition on $\sigma$. b: fixed boundary
      condition on $\sigma$.}}}
\end{figure}

A priori, it does not seem possible to use the manifestly conformally
invariant formulation (\ref{pert}) and the Neumann condition
(\ref{Neu}), which is the natural boundary condition at the UV fixed
point of the Coulomb gas, to describe the conformal boundary
conditions of the minimal models we are interested in.

\section{Boundary terms}

\subsection{Definition and usage}

A solution to this dilemma is the insertion of one or more of the
following boundary operators into the correlation functions
\begin{eqnarray} 
   B_{1,2}(x_0)\ &:=&\ \lim_{\delta\to0}\,(2\delta)^{2\,\Delta_{1,2}}
      \,V_{\alpha_{1,2}}(x_0+\ii\delta)\,V_{\alpha_{1,2}}(x_0-\ii\delta)
      \ ,\nonumber\\
   B_{2,1}(x_0)\ &:=&\ \lim_{\delta\to0}\,(2\delta)^{2\,\Delta_{2,1}}
      \,V_{\alpha_{2,1}}(x_0+\ii\delta)\,V_{\alpha_{2,1}}(x_0-\ii\delta)
      \ , \label{bot}
\end{eqnarray}
and to define the correlation function for a particular boundary
condition by
\begin{eqnarray}
   \bra X \ket^{\rm\,b.c.}\ &=&\ f_0^{\rm\,b.c.} \bra X \ket^{\rm Neumann}
      \ +\ f_1^{\rm\,b.c.} \bra B_{1,2}(x_0)\,X \ket^{\rm Neumann}\ + 
      \nonumber\\
   &&\ +\ f_2^{\rm\,b.c.} \bra B_{1,2}(x_0)\,B_{1,2}(x_1)\,X 
      \ket^{\rm Neumann}\ +\ \dots\ \label{bcexp}
\end{eqnarray}
The point $x_0$ is an arbitrary point on the boundary. The evaluation
of correlators with these insertions is defined by first balancing the
charges with the help of screening operators, then transforming the
two-dimensional integrals into contour integrals, and finally taking
the limit (or limites) of $\delta\to0$. In the following, we will
restrict our discussion to $B_{1,2}$. Equivalent statements are true
for $B_{2,1}$.

Selecting a preferred point, $x_0$, (or even several $x_i$) may seem
unnatural, but this will correspond to the points at which a contour
crosses the boundary, and the result will be independent of $x_0$.
Observe as well that $B_{1,2}$ has dimension 0. Suitably screened, it
can be viewed as the product of two $\sigma$ operators projected onto
the identity. Thus in the full-plane description, the remaining 
operators will be trivial.

The insertion of a $B_{1,2}$ generically causes correlation functions
to vanish because of the limit $\delta\to0$: the operator product
expansion of the two vertex operators contributes with a leading
factor $\delta^{2\alpha_{1,2}^2}$, the total exponent of $\delta$ is
therefore 
\begin{equation}
   2\,(\Delta_{1,2}+\alpha_{1,2}^2)\ > 0\ .
\end{equation}
However, the screening operator $V_{\alpha_-}$ in the upper half plane
has to encircle $V_{\alpha_{1,2}} (x_0+\ii\delta)$ on a Pochhammer
contour which lies entirely in the upper half plane, while its mirror
image has to go around $V_{\alpha_{1,2}} (x_0-\ii\delta)$ with a
contour in the lower half plane. During their trip along the
Pochhammer contour, these screening operators will pass momentarily,
i.e.~for a distance $\sim2\delta$, through a $\delta$-neighbourhood
around $x_0$. The leading term of their operator product expansion
with the other screeners and the two vertex operators in this
neighbourhood will produce additional factors of
$\delta^{2\alpha_i\alpha_j}$. Since the exponents of these factors are
negative, we have to carefully calculate the total exponent before
sending $\delta\to0$.

If there are $p$ screeners $V_{\alpha_+}$ and $q$ screeners
$V_{\alpha_-}$ in a $\delta$-neighbourhood of $x_0$, these factors are
\begin{equation} \label{brace}
   \underbrace{\delta^{2\Delta_{1,2}}\ \delta^{2\alpha_{1,2}^2}}
      _{\hbox{\tiny from $B_{1,2}$}}
   \underbrace{\delta^{p+q}}_{\vbox{\baselineskip7pt \hbox{\tiny\ \ 
      length of} \hbox{\tiny contribution}}}
   \underbrace{\delta^{4(p\alpha_++q\alpha_-)\alpha_{1,2}}}
      _{\hbox{\tiny screeners with $B_{1,2}$}}\ \ \,
   \underbrace{\delta^{2\frac{p(p-1)}2\alpha_+^2}\ 
      \delta^{2\frac{q(q-1)}2\alpha_-^2}\ \delta^{2pq\alpha_+\alpha_-}}
      _{\hbox{\tiny screeners with screeners}}\ .
\end{equation}
The total exponent is
\begin{equation} 
   [p\,\alpha_+ + (q-1)\,\alpha_-]^2\ \geq\ 0\ ,
\end{equation}
and vanishes if and only if $p=0$ and $q=1$. The corresponding result
for the insertion $B_{2,1}$ is: $p=1$, $q=0$. Note that the general
solution, $p=rM$ and $q=1+r(M+1)$, only needs to be considered for
$r=0$\,: for BRST reasons, a collection of $M$ screeners
$V_{\alpha_+}$ and $M+1$ screeners $V_{\alpha_-}$ vanishes when
applied on any vertex operator \cite{fel}.

Therefore, taking the limit $\delta\to0$ leaves only a non-vanishing
contribution if exactly one of the screeners around the two vertex
operators is trapped in the neighbourhood. The operator which remains
at $x_0$ is an uncharged identity operator. Having a contour coming
from a point $z$ in the upper half plane to $x_0$ with only an
identity operator at $x_0$ and another contour from there to a point
$\bar w$ in the lower half plane, is however equivalent to a single
contour joining $z$ and $\bar w$:
\begin{eqnarray} 
   &&k\,\int_z^{x_0}\!\d s\ \int_{\bar w}^{x_0}\!\d t\ \bra B_{1,2}(x_0) 
      V_{\alpha_-}(s) V_{\alpha_-}(t)\ X \ket\ =\nonumber\\
   &&=\ k\,\lim_{\delta\to0}\,(2\delta)^{\frac18}\!\!
      \int_z^{x_0+\ii\delta}\!\!\!\d s\!\!\int_{\bar w}^{x_0-\ii\delta}
      \!\!\!\d t\,\bra V_{\alpha_{1,2}}(x_0+\ii\delta) V_{\alpha_{1,2}}
      (x_0-\ii\delta) V_{\alpha_-}(s) V_{\alpha_-}(t)\ X \ket \nonumber\\
   &&=\ k\,\Big\{\int_z^{x_0}\!\!\d s\ \bra \eins(x_0)\ V_{\alpha_-}(s)\ 
      X \ket \cdot \nonumber\\
   &&\qquad\quad \cdot\ \lim_{\delta\to0} \Big( (2\delta)^{\frac18}
      \int_{\bar w}^{x_0-\ii\delta}\!\!\!\d t\ \bra V_{\alpha_{1,2}}
      (x_0+\ii\delta) V_{\alpha_{1,2}}(x_0-\ii\delta) V_{\alpha_-}(t) \ket
      + {\cal O}(\delta) \Big) + \nonumber\\
   &&\quad\ + \int_{\bar w}^{x_0}\!\d t\ \bra \eins(x_0)\ 
      V_{\alpha_-}(t)\ X \ket \cdot \nonumber\\
   &&\qquad\quad \cdot\ \lim_{\delta\to0} \Big( (2\delta)^{\frac18}
      \int_z^{x_0+\ii\delta}\!\!\!\d s\ \bra V_{\alpha_{1,2}}(x_0+\ii\delta) 
      V_{\alpha_{1,2}}(x_0-\ii\delta) V_{\alpha_-}(s)\ket 
      + {\cal O}(\delta) \Big) \Big\} \nonumber\\
   &&=\ \ii\ (-1)^\frac14\,k\,\coe1{\sqrt{2}}B(\coe14,\coe14)\,
      \int_z^{\bar w}\!\d s\ \bra V_{\alpha_-}(s) X \ket\ . \label{ident}
\end{eqnarray} 
In the last step, all powers of $\delta$ (which cancel each other)
were extracted, and it was used that the inner integral is an
incomplete Beta function $B_x(\coe14,\coe14)$ at $x = \coe{\bar
w}{2\ii\delta} \to -\infty$ which produces the additional
prefactors. Note that the final integral does obviously not depend on
the point $x_0$.

Therefore, the insertions $B_{1,2}$ and $B_{2,1}$ produce
boundary-crossing contours which were missing in the previous
section. The need of such contours is not restricted to two-point
functions. The conformal blocks without boundary-crossing contours are
too restricted to be the blocks of $N$-point functions for generic
conformal boundary conditions of the minimal model.

\subsection{Generalisation to $B_{n,m}$}

One can as well expect the appearance of generalised insertions
\begin{equation} 
   B_{n,m}(x_0)\ :=\ \lim_{\delta\to0} (2\delta)^{2\,\Delta_{n,m}}
   V_{\alpha_{n,m}}(x_0+\ii\delta)V_{\alpha_{n,m}}(x_0-\ii\delta)\ .
\end{equation}
These operators need $n-1$ screening operators $V_{\alpha_+}$ and
$m-1$ screening operators $V_{\alpha_-}$ encircling $V_{\alpha_{n,m}}
(x_0+\ii\delta)$ on Pochhammer contours which lie entirely in the
upper half plane, and the same amount of screeners around
$V_{\alpha_{n,m}} (x_0-\ii\delta)$ with contours in the lower half
plane. Similar to (\ref{brace}), $p$ screeners $V_{\alpha_+}$ and $q$ 
screeners $V_{\alpha_-}$ produce an exponent of $\delta$
\begin{equation} 
   [(n-p-1)\alpha_+ + (m-q-1)\alpha_-]^2\ \geq\ 0\ ,
\end{equation}
what vanishes if and only if $p=n-1$ and $q=m-1$. In the limit
$\delta\to0$, again only terms with exactly half of the screeners
trapped in the neighbourhood will survive, which all leave an identity
operator at $x_0$, together with several boundary crossing contours.

Although we will see that the $B_{n,m}$ could be a quite useful link
to Cardy's boundary states (see (\ref{carb}) or \cite{carnine}), we
wish to argue that the expressions $B_{1,2}$ and $B_{2,1}$ defined in
(\ref{bot}) are the fundamental objects, and that, by combining them
in the sense of quantum group representations \cite{gosi}, the general
$B_{n,m}$ are built up. One should however bear in mind that the
expansion (\ref{bcexp}) could equally well be written as
\begin{eqnarray}
   \bra X \ket^{\rm\,b.c.}\ &=&\ g_0^{\rm\,b.c.} \bra X \ket^{\rm Neumann}
      \ +\ g_1^{\rm\,b.c.} \bra B_{1,2}(x_0)\,X \ket^{\rm Neumann}\ + 
      \nonumber\\
   &&\ +\ g_2^{\rm\,b.c.} \bra B_{1,3}(x_0)\,X \ket^{\rm Neumann}\ +\ 
      \dots\ . \label{bcexp2}
\end{eqnarray}
As an example, we want to show how an insertion $B_{1,2} (x_0) B_{1,2}
(x_1)$ splits into a linear combination of $B_{1,1}(x_0)$ and
$B_{1,3}(x_0)$, if the limit $x_1\to x_0$ is taken before the limites
$\delta,\delta' \to0$. According to the discussion in section 4, it
would seem natural that in a basis of conformal blocks, the contours
in Figures 4a and 4b are the only contributions of $x_0$ and
$x_1$. However, the Dotsenko integrals between two singularities are
special Pochhammer contours of trefoil type $(1,1,N-2)$. From the
theory of multiple hypergeometric functions \cite{erd}, it is known
that these functions do not suffice to describe the general solution
of the differential equation. For $N\geq4$, one has as well to use
contours of types $(1,2,N-3)$, $(2,2,N-4)$, etc., which correspond to
generalised Horn functions. This means that we have to consider 
contours as in Figure 4c, too.

\begin{figure}[h]
\fig{5.1in}{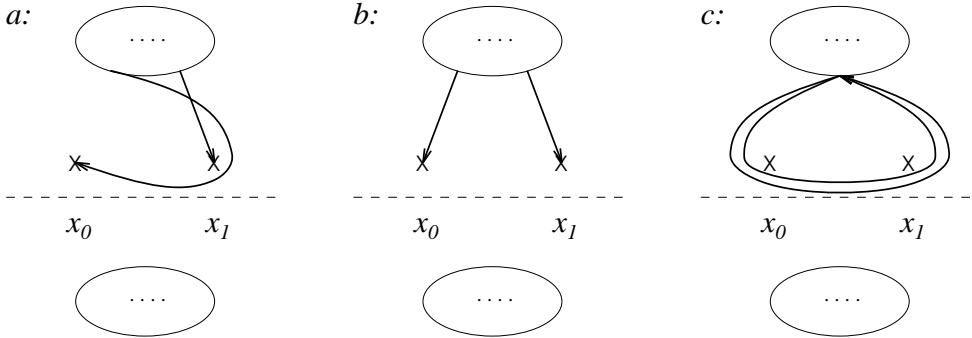}
\caption{Possible contours, the ellipses indicate other insertions}
\end{figure}

In the limit $x_1\to x_0$, the contours of Figures 4a and 4b join to
single contours and an identity operator as in (\ref{ident}). In
Figure 4c however, the two $V_{\alpha_{1,2}}$ vertex operators join to
$(x_1-x_0) ^{\frac38} V_{\alpha_{1,3}}$. Together with its mirror
image and the factor $\delta^{\frac18} \delta'^{\frac18}$, this is the
desired $B_{1,3}$. In the limit $x_1\to x_0$, the insertion
$B_{1,2}\,B_{1,2}$ contributes therefore as a linear combination of
$B_{1,1}$ and $B_{1,3}$.

Equivalently, it is seen that a triple contour around three $V_{1,2}$
operators at the boundary leaves a $B_{1,4}$ operator in the limit
$x_1\to x_0$, $x_2\to x_0$. In the Ising model, the $V_{1,4}$ vertex
operator is however $V_{1,4} = (Q V_{-2,0})$, where $Q$ is the BRST
operator, and is hence trivial \cite{fel,gosi}. Thus, there is a
truncation in the combination of $B_{1,2}$ insertions, what strongly
suggests that the $B_{n,m}$ are the highest weight vectors of multiple
tensor products of a representation of the quantum group $SU_q(2)$
with itself. This is true for generic $M$, using $q=\e^{\frac{2\pi\ii}
{2M}}$. The contributing correlators are in the $m=0$ state, which is
reached from the highest weight vector by application of an
appropriate amount of screening (i.e.~lowering) operators \cite{gosi}.

The truncation in the tensoring of the $B_{1,2}$ operators is a
consequence of the well-understood quantum-group structure of vertex
operators. It is however important for our situation, since it
guarantees that there are only finitely many terms we have to sum over
in (\ref{bcexp}) resp.~(\ref{bcexp2}) to get the conformal block of a
generic conformal boundary condition. In the Ising model for example,
there are only three terms contributing to get the conformal blocks of
the free and fixed boundary conditions.

\subsection{Boundary states and Virasoro eigenvalues}

Denote the boundary state without any $B_{1,2}$ insertions as $\vert
0,0 \ket$. The first zero stands for the $L_0$-eigenvalue
\begin{equation} \label{lzero}
   \int_{-\infty}^\infty\!\d x\,T(x)\, \vert0,0\ket = 0\ ,
\end{equation}
and the second zero is the $SU_q(2)$ label. Then the insertion of one
$B_{1,2}$ (and appropriate screeners) gives a state $\vert 1 \ket =
\sum_m c_m\vert\coe1{16},m\ket$. The insertion of two $B_{1,2}$ leads
with the help of the representation theory of the quantum group
$SU_q(2)$ to a linear combination of states $\vert 2 \ket = d \vert
0,0 \ket + \sum d_m\vert \coe12,m \ket$. Insertions of any odd (even)
number of $B_{1,2}$ give linear combinations of the form $\vert 1
\ket$ (resp.~$\vert 2 \ket$), with different coefficients. Note that
in the quantum group interpretation of the last subsection, the
$B_{n,m}$ are highest weight vectors which still have to be lowered by
screening operators.

Using (\ref{lzero}) and splitting the contour into the vanishing
integral on the real axis and a circle around $x_0 + \ii \delta$,
resp.~$x_0 - \ii\delta$, one immediately can verify that the first
label is both the $L_0$ and $\bar L_0$ eigenvalue. 

\subsection{Discussion for the Ising model}

In the next two subsections, we want to investigate the insertion of
$B_{1,2}$ terms in two examples. Especially, we would like to see how
much information we can gain on the coefficients $f_i^{\rm \,b.c.}$ in
(\ref{bcexp}).

From the remarks on truncation at the end of subsection 5.2, we
already know that for the Ising model only the first three of the
$f_i$ can be non-vanishing. Observe that for an odd number $N$ the
$N$-point function of the $\sigma$ operator vanishes for the free
boundary condition:
\begin{equation} 
   \bra \sigma(z,\bar z) \ket^{\rm free}\,=\,0\ ,\quad
   \bra\sigma(z_1,\bar z_1)\sigma(z_2,\bar z_2)\sigma(z_3,\bar z_3)
   \ket^{\rm free}\,=\,0\ ,\quad\dots
\end{equation}
This can be obtained in the Coulomb-gas picture by requiring that the
free boundary condition for $\sigma$ corresponds to a linear
combination of terms with even numbers of $B_{1,2}$ insertions only
(i.e.~$f_1^{\rm \,free}=0$). For an odd number of $V_{\alpha_{1,2}}$
in the upper half plane, the charges cannot be balanced by the
insertion of screeners, and all odd-point functions of the
$V_{\alpha_{1,2}}$ operator, resp.~$\sigma$, vanish. As it should be,
the $\Zset_2$ degeneracy for the free boundary condition appears as
well: all non-vanishing correlators are invariant under
$\sigma\to-\sigma$.

The fixed boundary condition, on the other hand, requires a
contribution of a single $B_{1,2}$ insertion for the one-point
function of $\sigma$. It should carry the sign of the external
magnetic field, which in the conformal limit is $h\to\pm\infty$. Since
the charges in the other to terms in (\ref{bcexp}) can not be
balanced, the one-point function would then read
\begin{eqnarray} 
   &&\bra\sigma\ket^{\rm fixed} = \nonumber\\
   &&=\ f_1^{\rm fixed}\ \lim_{\delta\to0}\ (2\delta)^{\frac18}\ \bra 
      V_{\alpha_{1,2}}(x_0+\ii\delta)V_{\alpha_{1,2}}(x_0-\ii\delta) 
      V_{\alpha_{1,2}}(z)V_{\alpha_{1,2}}(\bar z) 
      \ket_{{\cal A}_{\rm pert.}}\nonumber\\
   &&=\ k\,f_1^{\rm fixed}\ \lim_{\delta\to0}\ (2\delta)^{\frac18}
      \int_{x_0+\ii\delta}^{z}\!\d w \int_{x_0-\ii\delta}^{\bar z}
      \!\d\bar w \cdot \nonumber\\
   &&\quad\ \cdot\ \bra V_{\alpha_-}(w) V_{\alpha_-}(\bar w) 
      V_{\alpha_{1,2}}(x_0+\ii\delta)V_{\alpha_{1,2}}(x_0-\ii\delta) 
      V_{\alpha_{1,2}}(z) V_{\alpha_{1,2}}(\bar z) \ket^{\rm Neumann}
      \nonumber\\
   &&=\ \coe1{\sqrt{2}}\,B(\coe14,\coe14)\,k\,f_1^{\rm fixed}\ \cdot 
      \int_{z}^{\bar z}\!\d w\ \bra V_{\alpha_-}(w) V_{\alpha_{1,2}}(z) 
      V_{\alpha_{1,2}}(\bar z) \ket \nonumber\\
   &&=\ \coe1{\sqrt{2}}\,B(\coe14,\coe14)^2\,k\,f_1^{\rm fixed}\ \cdot\ 
      (z-\bar z)^{-\frac18}\ ,
\end{eqnarray}
where we have used (\ref{ident}). By comparing this result with
\cite{cale}\ one gets for the coefficients $k\,f_1^{\rm
fixed}\,B(\coe14,\coe14)^2\ =\ 2^{\frac34}$. A single $B_{1,2}$ can
however not be the full story for the fixed boundary condition. If
there were no contribution of an insertion of a pair of $B_{1,2}$ or
of the Neumann boundary condition, the two-point function of $\sigma$
would vanish. Recalling the conformal blocks (\ref{Cardy}) and
(\ref{fixed}), as well as the corresponding Figures 3a and 3b, one
knows that there has to be a contribution of $B_{1,2}\, B_{1,2}$ for
free and fixed boundary conditions. But a correlator with four
insertions in each half plane has five independent cross ratios and 32
conformal blocks. This makes it quite difficult to use the techniques
described in the Appendix to determine with which conformal block one
ended up. However, there is an easier way of gaining further insight:

Consider the one-point function of $\varepsilon$ in the expansion
(\ref{bcexp2}) for which as well only the first three terms
contribute. Only for the $B_{1,3}$-insertion, the charges can be
balanced, and the other two terms do not contribute. From the sign
change under the duality transform, one gets $g_2^{\rm fixed} =
-g_2^{\rm free}$. Now recall that equations (\ref{fixed}) and
(\ref{Cardy}) are the sum, respectively the difference, of the two
conformal blocks in which the $\eins$- and $\varepsilon$-channel
propagate through the boundary. Hence, $g_0^{\rm fixed}$ is $g_0^{\rm
free}$. With the help of the quantum group Clebsch-Gordon coefficients
the $g_i$ can be transformed into $f_i$.

This can be compared to Cardy's result in \cite{carnine}. He
introduces boundary states $\vert j \ket = \sum \vert j,N \ket \otimes
U \overline{\vert j,N \ket}$ which fulfill all the boundary conditions
required by the underlying $W$-algebra and are $W$-descendents of the
highest weight state $\vert j,0 \ket$ with eigenvalue $j$ under the
action of $L_0 \equiv \bar L_0$\footnote{These states should not be
confused with Cardy's $\widetilde{\vert j \ket}$: along an infinitely
long strip only $\Phi_j$ propagates for the boundary states
$\widetilde{\bra 0 \vert}$ and $\widetilde{\vert j \ket}$. Cardy maps
the strip later to a half plane, introducing boundary-condition
changing operators at the origin. Our boundary is equivalent to one
side of the strip, the other one being moved to an infinite distance,
we will not consider boundary-condition changing operators in this
paper.}. 

Cardy identifies the states corresponding to the conformal boundary
conditions of the Ising model as
\begin{eqnarray} 
   \vert f \ket\ &=&\ \vert 0 \ket - \vert \varepsilon \ket
      \ ,\nonumber\\
   \vert \pm \ket\ &=&\ \coe1{\sqrt{2}}\vert0\ket + \coe1{\sqrt{2}} 
      \vert \varepsilon \ket \pm \coe1{\root4\of2}\vert \sigma \ket
      \ . \label{carb}
\end{eqnarray}
Our results hence suggest that we should interpret (\ref{carb}) such
that we have to subtract the conformal block without boundary
insertions and the the one with a $B_{1,3}$-insertion from each other
to get the conformal block of the free boundary condition. The fixed
boundary condition equivalently would be given as a linear combination
of no insertion, a $B_{1,2}$-insertion and a $B_{1,3}$-insertion.

In this interpretation, the odd-point functions of the $\sigma$
operator vanish for free boundary conditions. The one-point function
carries the sign of the external magnetic for the fixed boundary
conditions. The sign change in front of the $I_2$-term in equations
(\ref{Cardy}) and (\ref{fixed}) is explained, as well the sign change
under the duality transform for $\bra \varepsilon \ket$\footnote{The
factor of $\sqrt{2}$ could be explained by a different normalisation
in (\ref{carb}).} \cite{cale}.

The similarity between (\ref{bcexp}) and (\ref{bcexp2}) with the
coefficients just derived, and (\ref{carb}) has to be used with
caution, though. Cardy's states are an infinite sum over
$W$-descendents, including the Virasoro descendents which have shifted
$L_0$ eigenvalues, and are furthermore not normalisable.

\subsection{A further example: the 3-state Potts model}

For the 3-state Potts model (the minimal Model at $M=5$), there are
branch cuts of order five for a screener going around an insertion
point. For this example, we do not have an as rigorous proof as for
the Ising model that the two-point function for the Neumann boundary
condition is the conformal block given by the double integral of
Figure 2a\footnote{Nevertheless, we still think that the arguments
given outside the proof are strong enough for this to be true.}. This
means especially that we cannot say anything about the relative
coefficient in front of the two versions of the integral, but that it
is given by a sum of phase factors of order five.

Consider the $\varepsilon$ operator (with conformal weight
$\Delta_{2,1} = \frac25$).  Its two-point function for the Neumann
boundary condition is given by the hypergeometric function
\begin{equation}
   \bra \varepsilon \varepsilon \ket^{\rm Neumann}\ 
   =\ {\rm c} \cdot (z_{13}z_{24})^{-\frac45} (1-\coe1\xi)^{-\frac45}
   \hyp21(-\coe85,-\coe15;-\coe25;\coe1\xi)\ .
\end{equation}
The free boundary condition of the 3-state Potts model is according to
\cite{carnine}\ given by the sum of $\vert 0 \ket$ and $\vert
\varepsilon \ket$. For the former however, we can not balance the
charges, and it has to vanish. Hence, the one-point function is given
as the conformal block given by a $V_{2,1}$ insertion with a $B_{2,1}$
at the boundary. We get
\begin{equation}
   \bra \varepsilon \ket^{\rm free}\ =\ {\rm const} \cdot (2y)^{-\frac45}\ ,
\end{equation}
what is in accordance with \cite{cale}.

\section{Outlook}

It would be a natural expectation that the boundary terms (\ref{bot})
can be derived by adding boundary terms to the action. Changing the 
action to 
\begin{eqnarray} 
   {\cal A}_{\rm free}\,\ &=&\ {\cal A} + B_{1,3}\ ,\nonumber\\
   {\cal A}_{\rm fixed}\ &=&\ {\cal A} \pm B_{1,2} \label{ffa}
\end{eqnarray}
would produce many of the qualitative features required above, as
e.g.~the vanishing of the odd-point functions or the sign of the
$\varepsilon$ one-point function. An exact calculation of the
appearing coefficients is however cumbersome, and is beyond the scope
of this paper. One of the difficulties lies in the fact that beside
the factor $k$ in front of the screening term in the action, there are
now additional free parameters in front of the $B_{n,m}$ in
(\ref{ffa}).

By a series expansion of the exponential $\e^{B_{1,2}(x_0)}$, with
multiple points $x_i$ for the higher order terms, the changed action
leads to an expansion of multiple insertions of $B_{1,2}$. In such an
expansion, one gets arbitrarily many $B_{1,2}$ insertions. However,
once there are more $B_{1,2}$ than other insertions, one is forced to
put screening contours around pairs of $V_{1,2}$ operators close to
the boundary. But such terms would as well appear in the partition
function, and one can use standard arguments of quantum field theory
to show that only the ``connected'' pieces contribute. As shown above,
these are only finitely many terms.

One can compare (\ref{ffa}) to the boundary integrals in \cite{ghza}\
where the off-critical Ising model is considered as an example. The
critical Ising model we considered so far is the massless limit of the
bosonisation of the free Majorana fermion field theory
\begin{equation} 
   S\ =\ \dint_{M} \d^2 z\ (\Psi\bar\partial\Psi - \bar\Psi\partial
      \bar\Psi + m \Psi\bar\Psi)\ .
\end{equation}
If one considers the off-critical theory with free boundary condition
on the half plane, one has to add a boundary term \cite{ghza}\
\begin{equation} \label{gzbt}
   S_{\rm h.p.}^{\rm free}\ =\ S + \coe12 \int_{\Rset} \d x\ 
      (\Psi\bar\Psi + a\partial_x a)\ .      
\end{equation}
Ghoshal and Zamolodchikov introduce $a$ as a fermionic boundary
field. The $\Psi\bar\Psi$ term is an $\varepsilon(z,\bar z)$ in the
critical theory. This corresponds to the suggested $B_{1,3}$ in
(\ref{ffa}).

The coupling to the external magnetic field $S_h = S^{\rm free} + h
\int \d x \sigma_{\rm B}(x)$, where $\sigma_B = \coe12(\Psi +
\bar\Psi)\,a$, disappears under the limit $m\to0$ unless
$h\to\pm\infty$, for which it leads to the fixed boundary
conditions. The boundary spin operator $\sigma_{\rm B}$ can easily be
related to the pair of $V_{\alpha_{1,2}}$ operators arising from
$B_{1,2}$ at the boundary.

The boundary terms (\ref{gzbt}) are still present after scaling the
theory to the conformal point. Rewritten via a one dimensional Stoke's
theorem -- the points where the branch cuts cross the real axis
forming the ``boundary'' of the real axis -- they might explain the
origin of $B_{1,2}$ and $B_{2,1}$. It even leaves the option to relate
them to the additional degree of freedom $a$ \cite{ghza}\ which, for
scaling reasons, must have dimension 0 like $B_{1,2}$. The
introduction of boundary spin operators was considered for more
general conformal models in \cite{war,lmss}.

Additions of boundary terms to the action, which are similarly to the
terms in (\ref{ffa}) no integrals, appear as well for Toda theories
\cite{bcdr,peza}. It remains unclear whether the $B_{n,m}$ can be
related to these boundary terms (recall that the Coulomb gas
considered here is the $A_1$ Toda theory). The boundary term in
\cite{bcdr,peza}\ produces for the ordinary Lie algebra $A_1$ the
``screeners'' of the right hand side of (\ref{obc}). It would be nice
to interpret this as the second conformal block (\ref{solNb}), which
has to be added to the Neumann term (\ref{solN}) to get the conformal
blocks of free and fixed boundary conditions. Since the higher minimal
models ($M>3$) have more than two conformal boundary conditions, but
are still described by $A_1$, this interpretation is however unlikely
to work.

\section{Conclusion}

The Neumann boundary condition on $\Phi$ is the natural boundary
condition for a generic non-affine Toda field theory viewed as the UV
limit of an affine theory. Treating the Coulomb-gas description of
minimal models as a perturbation of the free field theory by a
Liouville potential, the Neumann boundary condition leads to screening
contours which do not cross the boundary. On the other side, the
$N$-point functions for conformal boundary conditions of the minimal
model fall into conformal blocks which generically correspond to
boundary-crossing contours.

We introduced boundary insertions $B_{1,2}$ and $B_{2,1}$ (\ref{bot})
which can be described as composite operators of a vertex operator and
its mirror image. These insertions sew together contours from each
half plane to a boundary-crossing contour. Only an identity operator
remains at the point where this contour crosses the boundary, while
all other contributions vanish. It is argued that these insertions can
combine to general $B_{n,m}$. The appearing truncation follows from
the quantum group behaviour of the vertex operators, with $q =
\e^{\frac{2\pi\ii}{M}}$.

Although the calculational difficulty is increased by the additional
insertions, sensitive statements can be made in simple
situations. There is a close connection between Cardy's boundary
states (\ref{carb}) and the coefficients in (\ref{bcexp}).


\begin{ack}
I wish to express my gratitude to N.~Warner, who posed the question of
this paper, and who guided me to its final form. I'd like to thank
H.~Saleur for helpful comments. As well, I acknowledge the financial
support by the DAAD (Doktorandenstipendium HSPII/AUFE). This work was
supported in part by the DOE (DE-FG03-84ER40168).
\end{ack}


\appendix
\section{Derivation of the contour integrals from the 
two-dimensional integral}

\subsection{The differential equation}

Consider the $\sigma$-operator two-point function of the Ising model
on the half plane. Recall that this correlator, written as an analytic
four-point function has to fulfill the differential equation
\cite{bpz}\
\begin{equation} \label{deq} 
   D_1\,I\ =\ \Big[\,\coe43\,{\partial_1}^2 - \frac{\partial_2}{z_{12}}
   - \frac{\partial_3}{z_{13}} - \frac{\partial_4}{z_{14}} - \coe1{16} 
   (\frac1{{z_{12}}^2} + \frac1{{z_{13}}^2} + \frac1{{z_{14}}^2}) \Big] 
   I\ =\ 0
\end{equation}
and three similar equations with $z_2$, $z_3$ and $z_4$ taking the
special role. Plugging the integral representation (\ref{trick}) into
this differential equation one gets
\begin{equation} \label{deqA}
   D_1 \dint\!\d s\d t\ (\cdots)\ =\ \dint \!\d s\d t\ \Big( \partial_s 
   \frac{(\cdots)}{z_1-s} + \partial_t \frac{(\cdots)}{z_1-t} \Big)\ ,
\end{equation}
what obviously vanishes applying Stoke's theorem, if the $s$- and the
$t$-integra\-tions are along closed (i.e.~Pochhammer) contours.
Integrals between two pairs of singularities, plugged into
(\ref{deqA}), leave divergent expressions at the two end points, which
are not easily seen to cancel each other. However, a Dotsenko integral
between two singularities is up to a prefactor a Pochhammer integral
around the same pair of points\footnote{It is easy to see that the
branch cuts give weights $1+q+1-q^{-1}=2-2{\rm Re}\,q$ to the four
parts of the Pochhammer contours, and that the contributions in an
$\delta$-neighbourhood of the singularities are proportional to
$\delta \cdot \delta^{-\frac34} \sim \delta^{\frac14}\to
0$.}. Therefore, double integrals with both contours going from one
singularity to another fulfill (\ref{deq}), too.

For the area integral $\tint \d^2 z\ (\dots)$, Stoke's theorem applied
to (\ref{deqA}) leads to integrals along the branch cuts which vanish
since the integration along the left and right sides of the cuts
cancel each other because of monodromy invariance of the integrand.
For the half plane, there are as well integrals along the real axis,
but these vanish due to the factor $(z-z^*)^\frac32$. Therefore, the
area integrals fulfill the differential equations both for the full
and the half plane, and hence they are a linear combination of the
conformal blocks.

\subsection{Evaluation of the two-dimensional integral}

We argued in section 4 that it is a priori impossible to get boundary
crossing contours by evaluating the two-dimensional integral $\tint
\d^2 z\ (\dots)$. A screener and its mirror image are on two different
half planes, hence after applying Stoke's theorem as in (\ref{trick})
one integration is performed along the boundary of one half plane (the
boundary includes the branch cuts) while the other integration goes
from a reference point\footnote{If one chose the reference point in
the same half plane as the first integration variable, one would not
get a contradiction to this statement. The integral from this
reference point to the complex conjugate of the first integration
variable can always be performed via a new reference point in the
other half plane. It hence splits into the integral which we consider
and an integral with fixed end points. The later can be moved out of
the first integral, which then vanishes for homotopy reasons.} in the
other half plane to the complex conjugate of the first integration
variable. Therefore each integration variable stays in its respective
half plane. Since there are no distinguished points on the boundary,
the contours have to run between the singular points in the bulk.

To show for a certain example whether this argumentation is indeed
correct, one can use the following tools: 

$a)$ One chooses a boundary of the upper half plane by carefully
defining the branch cuts of $(z_i-s)^{2\alpha_-^{\ }\alpha_{n,m}}$. 
The branch cuts for $(\bar z_i-t)^{2\alpha_-^{\ }\alpha_{n,m}}$ then
have to be the mirror image of the former cuts.

$b)$ The boundary can then be cut into pieces between two singularities,
pieces going to the boundary and boundary pieces\footnote{The
integrals along the half circle at infinity is easily shown to
vanish. In the compactified picture of a half sphere instead of the
half plane, this means just that there is nothing special about the
point $z=\infty$.}. 

$c)$ The $J$-terms, which are in the upper half plane along one of these
pieces and in the lower half plane from one end point to the complex
conjugate of the first integration variable, appear in pairs which
cancel each other. The $J$-terms along the boundary vanish for reality
reasons (see $f)$ and $g)$). 

$d)$ Integrals along closed loops not encircling any singular point
vanish if and only if the double integral is not a $J$-term.

$e)$ Carefully calculating the monodromy coefficients, one moves all
pieces to one side of the branch cuts. 

$f)$ Since vertex operators and their mirror images have the same
charges, the complex conjugate of a double integral is obtained by
integrating between the complex conjugates of the end points (paying
careful attention to the branch cuts).

$g)$ From the way we defined the integrand after (\ref{foot}), it is
clear that integrals $\tint \d^2 z\ (\dots)$ are real. Application of
Stoke's theorem via (\ref{trick}) yields an $\frac{-\ii}2$, hence the
sum of the double integrals has to be purely imaginary. We can
subtract the complex conjugate of all these terms and divide by 2. 

For the two-point function of the $\sigma$ operator in the Ising model
(and as well for the $\Phi_{2,1}$ operator of the tri-critical Ising
model $M=4$), the integrals going to the boundary have vanishing
overall factor, and one remains with $\tint \d^2 z\ (\dots) = 1 \cdot
\int_{z_1}^{z_3}\!\d s \int_{z_2}^{z_4}\!\d t (\dots)$. This proof
does not work as straightforward for other examples, where one has to
use nontrivial relations between different integrals going to the
boundary, and it is hence far from being general\footnote{Even if one
argued that there might be an exotic example for which the
two-dimensional integral might give a conformal block which
corresponds to boundary-crossing contours, this would be {\it one}
well-defined block and to derive the others one needs to introduce the
$B_{n,m}$.}. Especially, there might be a proportionality factor
different from 1.

From the remarks made at the beginning of this subsection, one
furthermore expects that the contours are symmetric under reflection
along the real axis, a symmetry which ensures monodromy invariance and
reality. This statement is obviously fulfilled for the case considered
above. It is true in general since the two-dimensional integral has to
yield a function which depends on the {\it real} cross ratios only.

\subsection{Asymptotic behaviour}

For large $R=z_{12} \approx z_{34} \gg z_{13} \approx z_{24}$, the
asymptotic behaviour of the integral $\int_{z_1}^{z_3}\!\d s
\int_{z_2}^{z_4}\!\d t (\dots)$ gives after the substitutions $s=z_1 -
z_{13}u$ and $t=z_2 - z_{24}v$ two separate integrals for $u$ and $v$
in ${\cal O}(R^0)$, which are easily identified as Euler's
Beta-functions. ${\cal O} (R^{-1})$ vanishes. The subleading order is
${\cal O}(R^{-2})$ what leads with $\coe1\xi \sim \frac{z_{13}
z_{24}}{R^2}$ to
\begin{equation} 
   I_{\rm Neumann}\ \sim\ (z_{13}z_{24})^{-\frac18}(B(\coe14,\coe14)^2 
   R^0 + {\cal O}(\coe1\xi))\ .
\end{equation}
Therefore, there is no subleading contribution of ${\cal O}((z_{13}
z_{24})^{\frac38})$, i.e. no contribution of (\ref{solNb}) and the
integral is identified as (\ref{solN}).



\begin{thebibliography}{9}

\bibitem{ghza}{S. Ghoshal and A. Zamolodchikov,
   {\it Boundary S-Matrix and Boundary State in Two-Dimensional 
      Integrable Quantum Field Theory},
   Int. Jour. Mod. Phys. {\bf A9} (1994) 3841-3885.}

\bibitem{war}{N. Warner,
   {\it Supersymmetry in Boundary Integrable Models},
   Nucl. Phys. {\bf B450} (1995) 663-694.}

\bibitem{aflu}{I. Affleck and A.W. Ludwig,
   {\it Universal Noninteger ``Ground-State Degeneracy'' in Critical 
      Quantum Systems},
   Phys. Rev. Lett. {\bf 67} (1991) 161-164;
   {\it Critical Theory of Overscreened Kondo Fixed Points},
   Nucl. Phys. {\bf B360} (1991) 641-696.}

\bibitem{fls}{P. Fendley, A. Ludwig and H. Saleur,
   {\it Exact Hall Conductance through Point Contacts in the 
      $\nu=\coe13$ Fractional Quantum Hall Effect},
   Phys. Rev. Lett.{\bf 74} (1995) 3005-3008.}

\bibitem{cofl}{F. Constantinescu and R. Flume,
   {\it Perturbation Theory around Two-Dimensional Critical Systems 
      through Holomorphic Decomposition},
   Jour. Phys. {\bf A23} (1990) 2971-2986.}

\bibitem{homa}{T.J. Hollowood and P. Mansfield,
   {\it Rational Conformal Field Theories at, and away from 
      Criticality as Toda Field Theories},
   Phys. Lett. {\bf 226B} (1989) 73-79.}

\bibitem{egya}{T. Eguchi and S.-K. Yang,
   {\it Deformations of Conformal Field Theories and Soliton 
      Equations},
   Phys. Lett. {\bf 224B} (1989) 373-378.}

\bibitem{dofaA}{V.S. Dotsenko and V.A. Fateev,
   {\it Conformal Algebra and Multipoint Correlation Functions in 
      2d Statistical Models},
   Nucl. Phys. {\bf B240} (1984) 312-348.}

\bibitem{fel}{G. Felder,
   {\it BRST Approach to Minimal Models},
   Nucl. Phys. {\bf B317} (1989) 215-236;
   erratum, Nucl. Phys. {\bf B324} (1989) 548.}

\bibitem{mat}{S.D. Mathur,
   {\it Quantum Kac-Moody Symmetry in Integrable Field Theories},
   Nucl. Phys. {\bf B369} (1992) 433-460.}

\bibitem{carfour}{J.L. Cardy,
   {\it Conformal Invariance and Surface Critical Behavior},
   Nucl. Phys. {\bf B240} (1984) 514-532.}

\bibitem{carnine}{J.L. Cardy,
   {\it Boundary Conditions, Fusion Rules and the Verlinde Formula},
   Nucl. Phys. {\bf B324} (1989) 581-596.}

\bibitem{beg}{T.W. Burkhardt, E. Eisenriegler and I. Guim,
   {\it Conformal Theory of Energy Correlations in the Semi-Infinite
      Two-Dimensional $O(N)$ Model},
   Nucl. Phys. {\bf B316} (1989) 559-572.}

\bibitem{buxu}{T.W. Burkhardt and T. Xue,
   {\it Conformal Invariance and Critical Systems with Mixed Boundary 
      Conditions},
   Nucl. Phys. {\bf B354} (1991) 653-665.}

\bibitem{bcdr}{P. Bowcock, E. Corrigan, P.E. Dorey and R.H. Rietdijk,
   {\it Classically Integrable Boundary Conditions for Affine Toda
      Field Theory},
   Nucl. Phys. {\bf B445} (1995) 469-500.}

\bibitem{peza}{S. Penati and D. Zanon,
   {\it Quantum Integrability in Two-Dimensional Systems with 
      Boundary},
   Phys. Lett. {\bf 358B} (1995) 63-72.}

\bibitem{cale}{J.L. Cardy and D.C. Lewellen,
   {\it Bulk and Boundary Operators in Conformal Field Theory},
   Phys. Lett. {\bf 259B} (1991) 274-278.}

\bibitem{col}{S. Coleman,
   {\it Quantum Sine-Gordon Equation as the Massive Thirring Model},
   Phys. Rev. {\bf D11} (1975) 2088-2097.} 

\bibitem{man}{P. Mansfield,
   {\it Light-Cone Quantisation of the Liouville and Toda Field 
      Theories},
   Nucl. Phys. {\bf B222} (1983) 419-445.}

\bibitem{car}{J.L. Cardy,
   {\it Conformal Invariance and Statistical Mechanics},
   in E. Br\'ezin and J. Zinn-Justin (eds.),
   {\it Champs, Cordes et Ph\'enom\`enes Critiques}, 
      Les Houches XLIX, 1988.}

\bibitem{ssw}{H. Saleur, S. Skorik and N.P. Warner,
   {\it The Boundary Sine-Gordon Theory: Classical and Semiclassical 
      Analysis},
   Nucl. Phys. {\bf B441} (1995) 421-436.}

\bibitem{dofaB}{V.S. Dotsenko and V.A. Fateev,
   {\it Four-Point Correlation Functions and the Operator Algebra in 
      2d Conformal Invariant Theories with Central Charge $c\leq1$},
   Nucl. Phys. {\bf B251} (1985) 691-734.}

\bibitem{bnz}{J. Bagger, D. Nemeschansky and J.-B. Zuber, 
   {\it Minimal Model Correlation Functions on the Torus},
   Phys. Lett. {\bf 216B} (1989) 320-324.}

\bibitem{gosi}{C. G\'omez and G. Sierra,
   {\it Quantum Group Meaning of the Coulomb Gas},
   Phys. Lett. {\bf 240B} (1990) 149-157;
   {\it The Quantum Symmetry of Rational Conformal Field Theories},
   Nucl. Phys. {\bf B352} (1991) 791-828.}

\bibitem{erd}{A. Erd\'elyi,
   {\it Hypergeometric Functions of two Variables},
   Acta Math. {\bf 83} (1950) 131-164.}

\bibitem{lmss}{A. LeClair, G. Mussardo, H. Saleur and S. Skorik,
   {\it Boundary Energy and Boundary States in Integrable Quantum 
      Field Theories},
   Nucl. Phys. {\bf B453} (1995) 581-618.}

\bibitem{bpz}{A.A. Belavin, A.M. Polyakov and A.B. Zamolodchikov,
   {\it Infinite Conformal Symmetry in Two-Dimensional Quantum Field 
      Theory},
   Nucl. Phys. {\bf B241} (1984) 333-380.}

\end{thebibliography}
\end{document}